**Virtual Gram staining of label-free bacteria using darkfield microscopy and deep learning**

Çağatay Işıl[1,2,3], Hatice Ceylan Koydemir[4,5], Merve Eryilmaz[1,2,3], Kevin de Haan[1,2,3], Nir Pillar[1,2,3], Koray Mentesoglu[1], Aras Firat Unal[1,2,3], Yair Rivenson[1,2,3], Sukantha Chandrasekaran[6], Omai B. Garner[6] and Aydogan Ozcan[1,2,3,*]

## Affiliations

[1]Electrical and Computer Engineering Department, University of California, Los Angeles, CA, 90095, USA

[2]Bioengineering Department, University of California, Los Angeles, CA, 90095, USA

[3]California NanoSystems Institute (CNSI), University of California, Los Angeles, CA, 90095, USA

[4]Department of Biomedical Engineering, Texas A&M University, College Station, TX, 77843, USA

[5]Center for Remote Health Technologies and Systems, Texas A&M Engineering Experiment Station, College Station, TX, 77843, USA

[6]Department of Pathology and Laboratory Medicine, David Geffen School of Medicine, University of California, Los Angeles, CA, 90095, USA

* ozcan@ucla.edu

## Abstract

Gram staining has been one of the most frequently used staining protocols in microbiology for over a century, utilized across various fields, including diagnostics, food safety, and environmental monitoring. Its manual procedures make it vulnerable to staining errors and artifacts due to, e.g., operator inexperience and chemical variations. Here, we introduce virtual Gram staining of label-free bacteria using a trained deep neural network that digitally transforms darkfield images of unstained bacteria into their Gram-stained equivalents matching brightfield image contrast. After a one-time training effort, the virtual Gram staining model processes an axial stack of darkfield microscopy images of label-free bacteria (never seen before) to rapidly generate Gram staining, bypassing several chemical steps involved in the conventional staining process. We demonstrated the success of the virtual Gram staining workflow on label-free bacteria samples containing *Escherichia coli* and *Listeria innocua* by quantifying the staining accuracy of the virtual Gram staining model and comparing the chromatic and morphological features of the virtually stained bacteria against their chemically stained counterparts. This virtual bacteria staining framework effectively bypasses the traditional Gram staining protocol and its challenges, including stain standardization, operator errors, and sensitivity to chemical variations.

## Introduction

Gram staining, one of the fundamental techniques frequently used in microbiology for the past century, enables cost-effective classification of bacterial species into Gram-positive and Gram-negative groups based on their cell wall structure[1–4]. This differentiation has been crucial in many applications across diagnostics, food safety, and environmental monitoring, demonstrating its broad utility in various fields[5,6]. In clinical settings, Gram staining serves as a critical first step in identifying pathogenic bacteria in various bodily fluids such as cerebrospinal fluid (CSF), blood, and sputum[7]. By providing rapid results, Gram staining enables timely diagnosis of infections, particularly in cases of pneumonia, meningitis, and sepsis, which can be potentially life-threatening[8–10]. Bacterial differentiation based on morphology and cell wall composition through Gram staining allows physicians to quickly identify potential infections in bodily fluids, guiding decisions for further testing and initial antibiotic therapy[11–13]. Beyond its diagnostic utility, researchers have investigated the intricate interactions between antimicrobial drugs and target bacteria by using Gram staining[14,15]. These insights are critical for developing new antibiotics and formulating strategies to mitigate resistance. As such, Gram staining plays an important role in ongoing efforts to advance our understanding of bacterial pathogens[16,17].

While Gram staining remains a cornerstone technique in microbiology, it is not without limitations that warrant consideration in modern laboratory settings. Its manual sample preparation and time-consuming steps present challenges in high-throughput environments, potentially delaying critical diagnostic decisions[18]. Moreover, the manual handling of chemicals poses safety risks to laboratory personnel and necessitates strict safety protocols. Another issue is its susceptibility to operator-dependent variations, which sets challenges for the consistency and reproducibility of the Gram staining. Factors such as the level of technician expertise, sample preparation quality, and subtle differences in reagent composition can lead to inconsistent outcomes. Some common issues include false Gram determination resulting from improper decolorization and substantial cell loss during the washing steps, which can skew the results and potentially impact clinical decisions[4,19,20]. These limitations highlight the need for better standardization in staining and the exploration of alternative or complementary techniques to enhance the reliability and efficiency of bacterial identification in clinical and research settings. Advancements in other measurement techniques, e.g., fluorescence in situ hybridization (FISH)[21,22], fluorescence-based flow cytometry[23], polymerase chain reaction (PCR)[24,25], and next-generation sequencing (NGS)[26] present various powerful opportunities for direct pathogen identification without staining. Nonetheless, these advanced methods often require complex pre-treatment procedures, costly reagents, and extended turnaround times. While some of the automated systems show promise in addressing standardization challenges for accurate and repeatable Gram staining, they also present significant equipment expenses[27], limiting their widespread use. Despite these innovations and ongoing research into alternative methods, Gram staining remains one of the first choices for the initial evaluation of bacteria in microbiology.

Independent of imaging and sensing of bacteria, chemical staining procedures in the histopathology field have recently seen massive advances through deep learning-enabled chemistry-free virtual staining approaches, offering new possibilities for rapid, cost-effective, and repeatable tissue analysis[28–30]. These virtual tissue staining methods employ a deep neural network (DNN) model trained on a diligently processed dataset of paired or unpaired images, created using stained and unstained tissue sections. Following a one-time training phase, a virtual staining DNN can computationally transform microscopic images of label-free tissue samples (e.g., autofluorescence images) into their stained counterparts, matching e.g., brightfield images of the same tissue samples after the staining process. The efficacy of this virtual histological staining framework has been demonstrated for various stains[29–45], including immunohistochemical (IHC) stains[46,47]. While these virtual tissue staining methods have been successfully demonstrated for various types of histopathology stains, their potential in microbiological staining of label-free bacteria remained unexplored.

Here, we present the first demonstration of virtual bacterial staining that computationally generates Gram-staining using microscopic images of label-free bacteria; see Fig. 1. This deep learning-enabled method achieves virtual Gram staining by transforming a 3D axial stack of darkfield images of unstained bacteria into their Gram-stained brightfield counterparts, using a conditional generative adversarial network (cGAN)[48,49]. Unlike autofluorescence-based microscopy, darkfield imaging avoids the photobleaching of label-free bacteria and maintains a consistent signal-to-noise ratio and image contrast during sample scanning. After training a virtual Gram staining model using a large number of well-aligned image pairs (captured using darkfield and brightfield microscopy), this in-silico bacteria staining framework eliminates the need for time-consuming and manual steps of conventional Gram staining and rapidly produces repeatable staining results. We demonstrated this virtual Gram staining approach on label-free bacterial smears involving *Escherichia coli* (*E. coli*) and *Listeria innocua* (*L. innocua*), which are Gram-negative (staining red to pink) and Gram-positive (staining blue to purple), respectively. Our analyses with diverse quantification metrics, including bacterium-level metrics, show that the virtual Gram staining model accurately transforms darkfield image stacks of unstained bacterial samples containing both *E. coli* and *L. innocua* into their Gram-stained brightfield equivalent images. We compared the chromatic and morphological features of virtually and chemically Gram-stained bacteria, demonstrating the capability of the presented method to accurately stain both Gram-positive and Gram-negative bacteria. Our comprehensive analyses validate the efficacy of the presented label-free bacteria staining framework, eliminating the need for chemical Gram staining along with its drawbacks, including a manual chemical staining process, standardization issues, and sensitivity to chemical variations. The core concepts of the virtual Gram staining approach presented in our work can potentially be adapted to other staining techniques used in microbiology, enabling real-time virtual staining of live bacteria, and also become part of automated systems[27,50,51].

## Results

Figure 1a illustrates the steps used in virtual Gram staining of label-free bacteria using deep learning compared to the traditional Gram staining process. To demonstrate this in-silico bacteria staining framework, a virtual Gram staining model based on the cGAN framework, as illustrated in Supplementary Fig. S1, was trained using well-aligned darkfield axial image stacks of label-free bacteria (covering an axial range of [-2:1:2 μm]) and the corresponding brightfield images of Gram-stained bacteria. These darkfield-brightfield image pairs (each with 1024 × 1024 pixels) include 200 image pairs containing only *E. coli*, another 200 image pairs with only *L. innocua*, and 750 image pairs featuring both *E. coli* and *L. innocua* (see the Methods section for details). The trained models were blindly tested using 1000 unique images that were never used before: 200 images containing both *E. coli* and *L. innocua* (*mixed test dataset*), 400 *E. coli*-only images, and 400 *L. innocua*-only images (*monoculture test datasets* including single species). In Figs. 1a-b, a virtually stained field-of-view (FOV) and the corresponding target, chemically stained brightfield image with their zoomed-in regions of interest (ROIs) are shown, demonstrating the microscopic-level staining accuracy of the virtual bacterial staining approach.

Figure 2 further compares the virtual Gram staining method against the conventional Gram staining. Darkfield axial image stacks of two test FOVs, covering an axial range of [-2:1:2 μm], were processed using the virtual and standard Gram staining techniques. The resulting images presenting virtually stained and chemically stained bacteria are depicted in Fig. 2a. In the right column, the corresponding accuracy map for each FOV illustrates the *true positive* ($TP$), *false negative* ($FN$), *false staining* ($FS$), and *bacteria hallucination* ($BH$) predictions for each detected bacterium; see the Methods section for details. $TP$ denotes an individual bacterium correctly stained (either Gram-negative or -positive) by the virtual staining model. $FN$ refers to an individual bacterium present in the chemically stained target images that the virtual staining model missed. $FS$ and $BH$ cases represent bacteria incorrectly stained and hallucinated by the model, respectively. These representative FOVs and zoomed-in ROIs in Fig. 2a demonstrate a very good visual agreement between the virtual and the conventional Gram staining methods. Additionally, Fig. 2b presents a table of quantitative metrics calculated across 200 images of the *mixed test dataset*. Our virtual staining framework achieves high Peak-Signal-to-Noise Ratio (PSNR) and Structural Similarity Index Measure (SSIM) values of 38.35 dB and 0.9685, respectively, further supporting its success. We also computed additional metrics, including precision (95.5%), recall (96.5%), F1-score (96%), $FS\ rate$ (~3%), and $BH\ rate$ (~1.5%) to demonstrate the high accuracy of the virtual Gram staining approach.

In addition to these, Figure 3 compares the chromatic and morphological features of bacteria in the virtually and chemically stained image FOVs. Four morphological features (i.e., surface area, diameter, circularity, and eccentricity) and four chromatic features (i.e., red, green, and blue color intensities as well as a red/blue intensity ratio) were analyzed for both the virtually and chemically stained bacteria. For this analysis, feature distributions of *E. coli* and *L. innocua* were computed using each segmented bacterium ($TP$ cases) in the target Gram-stained images and their corresponding virtually stained counterparts, resulting in a total of 33,455 bacteria from 200 images of the *mixed test dataset*. As demonstrated in Fig. 3, the distributions of various bacterial features from the virtually stained images closely match those of the chemically stained images. The difference between these distributions was quantified using the Hellinger distance (HD)[52,53], with the resulting values listed in Fig. 3 for each case (see the Methods section for details). The lower HD values that are observed along with the visual correlation of the feature distributions demonstrate the physical consistency of the virtual Gram staining approach. These results also show that the virtual Gram staining model accurately captures the inherent nature of the Gram staining process in terms of chromatic and morphological features, enabling accurate in-silico Gram staining.

Next, we compared the bacteria concentration levels inferred from virtually and chemically stained FOVs; see Fig. 4. We used test image patches created using the *mixed* and *monoculture test datasets* for this analysis. The number of bacteria in each virtually or chemically stained FOV was counted using a bacteria segmentation method (see Methods for details). These counts were split into groups of 9, and the count data from each group was aggregated to obtain the number of bacteria in each FOV of 0.1 mm × 0.1 mm. The resulting bacteria concentrations are illustrated in Fig. 4. To quantify the agreement between the concentration levels inferred using the virtual and chemical Gram staining methods, we calculated the coefficient of determination ($R^2$) and normalized root mean square error ($NRMSE$) between the predicted and ground truth bacterial concentrations[54]. For monocultures, the $R^2$ ($NRMSE$) values for *L. innocua*-only and *E. coli*-only cases were ~0.821 (~0.082) and ~0.968 (~0.102), respectively. For mixed cultures, the $R^2$ ($NRMSE$) values for *L. innocua* and *E. coli* were ~0.904 (~0.080) and ~0.846 (~0.111), respectively. These relatively small $NRMSE$ and high $R^2$ values demonstrate that our in-silico virtual staining approach is also accurate for inferring the concentration levels of Gram +/- bacteria.

Figure 5 further analyzes the potential impact of different size distributions of bacteria on training the virtual Gram staining model. As seen in Fig. 3, the size distributions (e.g. surface area, diameter) of the bacteria used in our results differ significantly. For instance, we computed the median surface areas of *E. coli* and *L. innocua* using 5 ground truth images selected to have minimal overlap of Gram-negative and Gram-positive bacteria. The resulting median surface areas are 0.892 $\mu m^2$ for *E. coli* and 0.317 $\mu m^2$ for *L. innocua*, illustrated with vertical dashed lines in Fig. 5. In this analysis, we examined whether this divergence in size distributions introduced any bias to the virtual Gram staining model. To assess this, we computed the accuracy of the virtual staining model in terms of precision and recall for similarly sized bacteria ($\rho \leq \text{surface area} \leq \rho + 0.1\ \mu m^2$, for each point ρ on the x-axis of Fig. 5) using the *mixed test dataset*. The precision and recall graph as a function of ρ, illustrated in Fig. 5, further supports the success of the virtual Gram staining approach: both the precision and recall values of Gram-positive and Gram-negative cases are greater than 0.83 for $\rho \geq 0.4\ \mu m^2$. The precision and recall values for Gram-negative bacteria are relatively lower for smaller ρ cases (e.g., $\rho = 0.2\ \mu m^2$). This could be due to possible color variations in the chemical Gram staining procedure rather than a limitation of the virtual staining model itself (see e.g., Supplementary Fig. S2); such chemical staining variations increase the numbers of $FP$ and $FN$ cases for Gram-negative bacteria with a relatively lower overall population in this size range. Despite the notable size differences between *E. coli* and *L. innocua*, our virtual Gram staining model demonstrates robust performance across a wide range of bacteria sizes. The consistently high precision and recall values for Gram-positive and Gram-negative bacteria that we observed in Fig. 5 indicate that the virtual staining model does not overfit to the size of the bacteria. Overall, these results and analyses underscore the success of the virtual Gram staining model in generalizing across different bacteria morphologies, highlighting its potential for broad applicability in microbiological analysis.

## Discussion

This manuscript demonstrates a virtual Gram staining approach using deep learning and darkfield microscopy. A trained virtual Gram staining model, based on the cGAN architecture, rapidly transforms an axial stack of label-free darkfield images of bacterial smears to match the corresponding brightfield images of Gram-stained bacteria. This technique can be rapidly performed on a consumer-grade computer, circumventing the manual and delicate steps involved in the traditional Gram staining process. After a one-time training effort, the presented virtual Gram staining method operates on darkfield images of label-free bacteria and digitally generates staining results with high consistency and repeatability. This contrasts with the conventional chemical Gram staining process, which is sensitive to human errors and chemical variations[4,19,20]. While sample quality and environmental factors can affect traditional Gram staining, the virtual bacterial staining approach is robust against these external sources of error and artifacts, which can cause problems in staining results, including color inconsistencies.

Darkfield imaging (including both in-focus and out-of-focus microscopic images) of label-free bacteria offers various benefits in our virtual staining framework, including high contrast and the ability to reveal fine bacterial structures along with their 3D optical scattering information[55–57]. The primary mechanism of conventional Gram staining relies on the thickness of the peptidoglycan polymer in the cell wall of each bacterium[6]. Darkfield microscopy captures the scattered light from the label-free bacterial cell walls, enabling differentiation between Gram-positive and Gram-negative bacteria and facilitating accurate virtual Gram staining. Compared to earlier works that used autofluorescence-based imaging, which is a commonly used optical modality for virtual staining of label-free tissue[30], darkfield microscopy avoids photobleaching of bacteria and maintains consistent signal-to-noise ratio (SNR) and signal intensity across the entire image FOV, which is crucial for label-free bacteria imaging[58–60]. These advantages of darkfield microscopy may also facilitate real-time virtual staining of live bacteria.

While darkfield input image stacks (axially spanning [-2:1:2 μm]) of label-free bacteria were used for training and testing the virtual Gram staining models reported so far, the presented approach could also function with different darkfield input stacks, e.g., [-1:1:1] μm or an in-focus image with [Z=0] μm – although with some compromise in performance. In Supplementary Fig. S3, we conducted an additional study to examine the accuracy of various virtual staining models trained on different darkfield input image stacks. We tested these designs using the *mixed test dataset* containing 200 images, each including both *E. coli* and *L. innocua*. This figure displays the overall precision, recall, and F1-score values for different sets of darkfield input images. This analysis not only demonstrates the effectiveness of including defocused darkfield image stacks compared to a single in-focus darkfield image as input for our deep learning-enabled staining approach but also reveals that a virtual bacterial staining model can achieve better performance with an increased number of defocused darkfield images in the input stack. This observation further suggests that the 3D scattering information of each bacterium captured through an axial stack of in-focus as well as out-of-focus darkfield images is essential for the success of our virtual Gram staining approach.

Although we successfully demonstrated the overall performance of our virtual Gram staining approach through our results and related analyses, several bacterium-level error sources also affect the accuracy of the trained models. As illustrated in Supplementary Fig. S2, some of these error sources include bacterial loss during the conventional Gram staining process, spatially obscured bacteria in label-free darkfield images, and color variations observed in the stained bacteria due to various factors such as over-decolorization and aging of individual bacteria[4,19,20]. These issues diminished the overall quality of the training image dataset. The same inaccuracies or imperfections also influenced the test datasets used for calculating the accuracy maps, ultimately leading to inflated error rates for the virtual Gram staining models since such chemical staining variations/artifacts increase the numbers of $FP$ and $FN$ cases. In addition to the error sources mentioned above, potential pixel-level inaccuracies in the segmentation model may have further contributed to some of our errors (see Supplementary Fig. S4 and the related Methods section).

In summary, we reported the first demonstration of virtual Gram staining that accurately transforms darkfield images of label-free bacteria into corresponding brightfield images presenting Gram-stained bacteria. This in-silico virtual staining technique, running on a consumer-grade computer, significantly streamlines the staining process, eliminating the manual and delicate nature of conventional staining procedures. Once trained, the system generates highly consistent and repeatable results from label-free darkfield images, circumventing the variability introduced by human error and chemical inconsistencies inherent in traditional Gram staining. Notably, this virtual bacterial staining approach demonstrates robustness against external factors such as sample quality and environmental conditions, which often compromise traditional staining outcomes. By addressing the limitations of conventional methods, virtual Gram staining represents a significant advancement in microbiology, offering a more repeatable, faster, cost-effective, and accessible tool for bacterial classification and analysis. The fundamental principles of the presented virtual bacterial staining can potentially be extended to other types of microbiological staining, including acid-fast and endospore staining[50,51].

## Methods

## Sample preparation

We used *L. innocua* (ATCC® 51742™) and *E. coli* (ATCC® 25922TM) as Gram-positive bacteria and Gram-negative bacteria, respectively. Following the instructions provided by the ATCC (American Type Culture Collection), bacteria were activated, and bacterial stock solutions were prepared and stored at -80 °C. After thawing the frozen bacterial stock solutions at room temperature, a loop of *L. innocua* was added to 10 mL of brain heart infusion broth, and a loop of E. coli culture was added to 10 mL of tryptic soy broth. Then, both mixtures were incubated at 30 °C overnight. After overnight incubation, the following experimental steps were repeated separately for *L. innocua* and *E. coli*.

Once the bacteria culture was ready, 1.0 mL of freshly prepared culture was added to 9.0 mL of sterilized reagent water. We added 1.0 mL of the suspension in a cuvette to determine the concentration of the bacteria in the suspension using a spectrophotometer (NanoDrop, Thermo Fisher). This step was crucial to ensure consistent bacterial concentration for spiking cerebrospinal fluid (CSF) (Innovative Research) and to accurately adjust the different ratios of *E. coli* and *L. innocua* in the CSF. Knowing the concentration of the prepared bacterial stock solutions allowed for the preparation of CSF suspensions spiked with *L. innocua* and/or *E. coli* at different ratios, i.e., 0:6, 3:3, 6:0, and concentrations. We aimed to have a final bacterial concentration of $6 \times 10^6$ CFU/mL in the CSF.

After preparing the bacteria-spiked CSF bacterial suspensions, 50 µL of each suspension was placed on glass slides and dried on a hot plate. Following drying, the slides were dipped in methanol and left for 5 minutes to fix the samples. After fixation, the slides were rinsed with water and dried gently using compressed air, prior to imaging using a darkfield microscope (see the Image acquisition sub-section).

**Gram staining procedure**

To stain the samples, the slides were first immersed in crystal violet solution for 1 minute, allowing the primary stain to penetrate the bacterial cells. The slides were then rinsed with water to remove excess stain. Next, the slides were dipped in Gram iodine solution for 1 minute. The iodine acts as a mordant, forming a complex with the crystal violet and trapping it within the cell walls. Following another rinse, a decolorization solution (5% ethanol in pure acetone) was gently applied for ~10 s. This step is crucial as it differentiates between Gram-positive and Gram-negative bacteria: Gram-positive bacteria retain the crystal violet-iodine complex and remain purple, while Gram-negative bacteria lose the complex and become colorless. After a water rinse step, the slides were immersed in Gram safranin solution for 15 s. Safranin serves as a counterstain, coloring the Gram-negative bacteria pink while the Gram-positive bacteria remain purple. Finally, after one more rinse and drying step, the slides were ready for brightfield imaging (ground truth).

**Image acquisition**

Darkfield axial image stacks of the bacterial smears with Z = [-2:1:2 µm] were captured using a standard inverted microscope (IX-83, Olympus) equipped with a 100× oil objective lens (0.6-1.3 NA, UPlanFL, Olympus) and a darkfield oil condenser (1.2-1.4 NA, U-DCW, Olympus). Each darkfield image was captured with a scientific complementary metal-oxide-semiconductor (sCMOS) image sensor (ORCA-flash4.0 V2, Hamamatsu Photonics), featuring a pixel size of 6.5 µm and an exposure time of ~20 ms. The image acquisition process was controlled by MetaMorph microscope automation software (version 7.10.161, Molecular Devices), which performed translational scanning and auto-focusing at each stage position. After the Gram staining procedure, brightfield images of the same bacterial smears were captured using the same 100× objective lens and a color camera (QImaging Retiga 4000R) with a pixel size of 7.4 µm × 7.4 µm and exposure time of 10 ms.

**Image preprocessing and pixel-level matching of darkfield and brightfield images**

We applied supervised learning using pixel-to-pixel matched darkfield-brightfield image pairs to train a virtual staining model. A critical step in this process is image registration to obtain these pixel-level aligned image pairs. The image registration pipeline is illustrated in Supplementary Fig. S5, which was implemented in MATLAB (version R2023a, MathWorks). Initially, darkfield images (pre-Gram staining) and the corresponding brightfield images (post-Gram staining) of the same bacterial smears were stitched into large FOVs. These FOVs were globally matched using a control point registration technique[61,62], as depicted in Supplementary Fig. S5a. In this technique, a few control points (e.g., 4-5) were selected, each representing approximately the same location in both the darkfield and brightfield FOVs. These points were used to estimate the parameters of an affine transformation matrix that was subsequently applied to the brightfield FOV to achieve a rough alignment with the corresponding darkfield FOV.

The second step of the image registration process started with cropping the roughly matched large FOVs into pairs of smaller image patches (1024 × 1024 pixels), as illustrated in Supplementary Fig. S5b. Due to optical aberrations, local movements, and morphological changes during the chemical Gram staining procedure, these cropped image pairs were not perfectly matched at the pixel level. To address this, the second step involved the local registration of these image patch pairs using pyramid elastic registration[63] and a proxy deep neural network[30,46]. First, the cropped image patches from the first step were used to train a proxy virtual staining network, transforming and matching the style of the darkfield images with the brightfield image patches. The network outputs and the corresponding brightfield patches were then used in the pyramid elastic registration method, which operates at multiple scales to hierarchically match the local features of sub-image blocks and elastically morph the brightfield image patches for finer image registration with respect to the corresponding darkfield images. This process was repeated until the desired pixel-level registration accuracy was achieved. Between these elastic registration cycles and at the final stage, correlation-based cleaning and manual cleaning of image pairs were performed to remove FOVs with staining artifacts and other discrepancies resulting from human-dependent chemical staining and sample preparation steps.

**Training and testing datasets**

The processed and well-aligned darkfield-brightfield image pairs (1024 × 1024 pixels) were used to create training, validation, and test datasets. These datasets include darkfield axial image stacks [-2:1:2 µm] of unstained FOVs and the corresponding brightfield images of the same FOVs after the chemical Gram staining. The training dataset consists of 200 images containing *E. coli*, another 200 images with *L. innocua*, and 750 images featuring both *E. coli* and *L. innocua*. The validation dataset includes 20 images containing *E. coli*, another 20 images with *L. innocua*, and 50 images featuring both *E. coli* and *L. innocua*. To measure the virtual Gram staining success of the trained models on mixed cultures, we used the *mixed test dataset*, including 200 images, each containing both *E. coli* and *L. innocua*, which were not seen by the model during training. Additionally, we tested the trained models on new FOVs containing single species using *monoculture test datasets*, including 400 images of only *E. coli* and 400 images of only *L. innocua*.

For the concentration analysis of the developed in-silico staining method in Fig. 4, image patches with 512 × 512 pixels were randomly cropped from the *mixed* and *monoculture test image datasets*. After this process, 1800 test image patches for the concentration analysis of mixed cultures and 3600 test image patches for the concentration analysis of each monoculture were used to test the concentration accuracy of the virtual Gram staining approach.

**Numerical evaluation metrics for virtually stained images**

Common detection metrics, including precision, recall, and F1-score, were utilized to evaluate the accuracy of the trained virtual staining models in addition to standard image quality metrics such as PSNR and SSIM[64,69]. Precision, recall, and F1-score are defined as follows:

$$Precision = \frac{N_{TP}}{N_{TP} + N_{FP}}$$

$$Recall = \frac{N_{TP}}{N_{TP} + N_{FN}}$$

$$F1 - score = \frac{2 \times Precision \times Recall}{Precision + Recall}$$

where $N_{TP}$ refers to the number of $TP$ bacteria that are correctly stained (either Gram-negative or Gram-positive) by the virtual staining model. $N_{FN}$ is the number of $FN$ bacteria present in the stained target images but missed by the trained neural network model. $N_{FP}$ denotes the number of $FP$ bacteria that are incorrectly stained or hallucinated by the virtual staining model. To analyze the effects of the hallucinated and incorrectly stained bacteria separately, we split $N_{FP} = N_{BH} + N_{FS}$ where $N_{BH}$ and $N_{FS}$ represent the numbers of hallucinated ($BH$) and incorrectly stained ($FS$) bacteria, respectively. We accordingly defined bacteria hallucination rate ($BH\ rate$) and false staining rate ($FS\ rate$) as follows:

$$BH\ rate = \frac{N_{BH}}{N_{TP} + N_{FP}}$$

$$FS\ rate = \frac{N_{FS}}{N_{TP} + N_{FP}}$$

In addition to these five detection-based metrics evaluating the accuracy of the developed virtual Gram staining method, PSNR and SSIM metrics were also utilized. PSNR value for a virtually stained image is computed as

$$PSNR = 10 \log_{10}\left(\frac{1}{MSE\{G(I_{in}), I_{tar}\}}\right).$$

SSIM value for a virtually stained image is computed using the standard MATLAB implementation with a maximum value of 1. These evaluation metrics were implemented in MATLAB (version R2023a, MathWorks).

For the concentration analysis in Fig. 4, $R^2$ and $NRMSE$ values were used to quantify the amount of concentration error between the images of virtually and chemically stained bacteria[54], which are defined as follows:

$$R^2 = 1 - \frac{\sum_i (gt_i - out_i)^2}{\sum_i (gt_i - \overline{gt})^2}$$

$$NRMSE = \frac{\sqrt{MSE\{gt, out\}}}{\overline{gt}}$$

where $gt_i$ and $out_i$ denote the bacteria concentration computed using each Gram-stained and virtually stained FOV covering 0.1 mm × 0.1 mm, respectively. $\overline{gt}$ is the mean bacteria concentration obtained from Gram-stained target images.

**Image segmentation method**

We utilized a combination of thresholding, clustering, and morphological operations for the segmentation of bacteria. This method ensures high accuracy and reliability in our analyses, enhancing the overall efficacy of our detection metrics. In our segmentation method, illustrated in Supplementary Fig. S6, RGB images are first converted to grayscale. These grayscale images are then processed using the global Otsu thresholding method[70] to create an initial segmentation map for bacteria, although this method is not accurate at the boundaries. The grayscale images are also processed by the local Otsu thresholding method with a patch size of 15 × 15 pixels, which provides more accurate segmentation near the boundaries of individual bacteria[71]. To improve the overall segmentation map accuracy, the K-means approach[72,73] with two clusters is applied to the pixel values of the RGB images to segment bacteria and background regions. Subsequently, a logical AND operation combines the segmentation maps from the local Otsu thresholding and the K-means clustering methods. This process is repeated to prevent convergence to a local minimum while monitoring the correlation coefficient between the segmentation maps of this combination and global Otsu thresholding with respect to a predefined threshold of 0.6.

After obtaining the resulting segmentation map corresponding to bacterial regions, several morphological operations, including erosion and dilation, are applied. Artifact removal is also performed using prior color and size distributions of bacteria[74]. Finally, the median color (RGB) of each individual bacterium is calculated using the overall segmentation map and the chemically/virtually Gram-stained images to assign a class (Gram-negative or Gram-positive) to each bacterium. A sample segmentation map for an FOV, including mixed bacteria culture and resulting Gram-positive and Gram-negative bacteria maps, are illustrated in Supplementary Fig. S6a. These computed bacteria maps were used in our workflow presented in Supplementary Fig. S6b to assign a classification outcome ($TP, FP\ (BH + FS), or\ FN$) for each segmented bacterium. Our image segmentation method and the corresponding workflow were implemented in MATLAB (version R2023a, MathWorks).

To evaluate the accuracy of our segmentation approach, it was compared against two human experts, as shown in Supplementary Fig. S4. Five brightfield images of Gram-stained bacteria, presented in Supplementary Fig. S4a, were provided to two human experts and processed by our method for counting *E. coli* and *L. innocua*. Supplementary Fig. S4b lists the bacteria counts for each image, along with the averages obtained by the human experts and our computational method. The mean absolute percentage error[75] of 5.19% and 4.12% (for *E. coli* and *L. innocua*, respectively) between the counts from our method and the average counts of the human experts calculated across these five images demonstrate comparable counting performance between our method and human experts. This result further validates the segmentation accuracy of the developed approach.

## Feature extraction of segmented bacteria and quantitative analysis of feature distributions

Figure 3 illustrates an analysis of feature distributions extracted from virtually and chemically stained images, along with their corresponding segmentation maps. A total of 33,455 segmented bacteria (only $TP$ cases) from 200 images of the *mixed test dataset* were used to compute these feature distributions using MATLAB (version R2023a, MathWorks). HD[52,53] was used to quantify the similarity between the feature distributions of bacteria from the network output (virtually stained) and the target (chemically stained) images. HD is defined as follows:

$$HD\big(p(x), q(x)\big) = \sqrt{1 - \sum_{x \in X} \sqrt{p(x)q(x)}}$$

where $p(x)$ and $q(x)$ are probability distributions of the features (approximated using extracted feature points). HD ranges between 0 (identical) and 1 (completely different).

**Supplementary Information** includes:
- Supplementary Figs. S1-S6
- Neural network architecture and training settings for virtual Gram staining

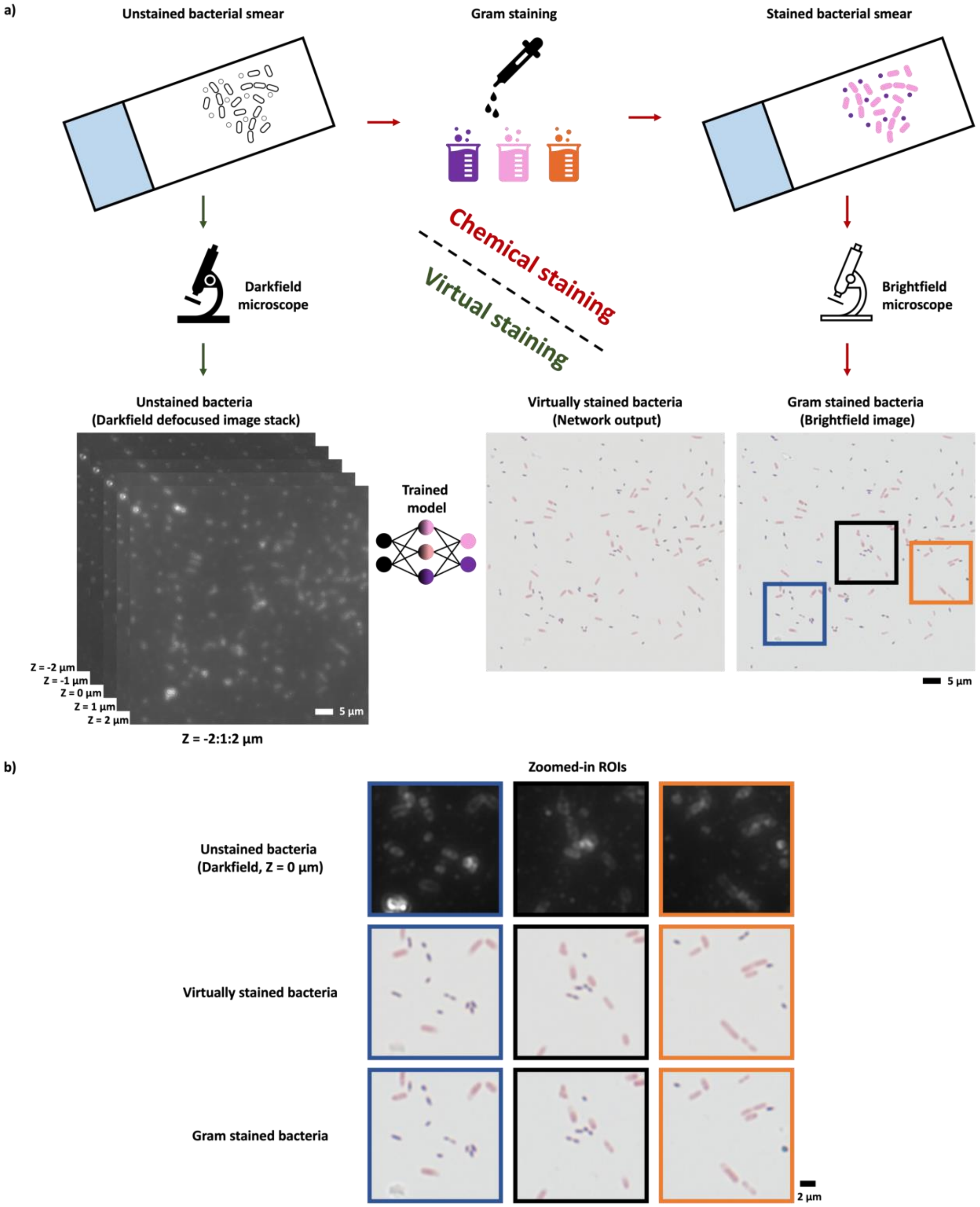

**Figure 1. Virtual Gram staining of unlabeled bacteria via deep learning.** (a) Conventional Gram staining is susceptible to human errors and involves manual sample processing in contrast to virtual Gram staining, which rapidly translates the darkfield axial image stack of label-free bacteria into brightfield equivalent images, matching their Gram-stained counterparts, circumventing chemical reagents and manual processing in the staining workflow. (b) Zoomed-in ROIs comparing the results of the virtual Gram staining with the conventional chemical staining approach.

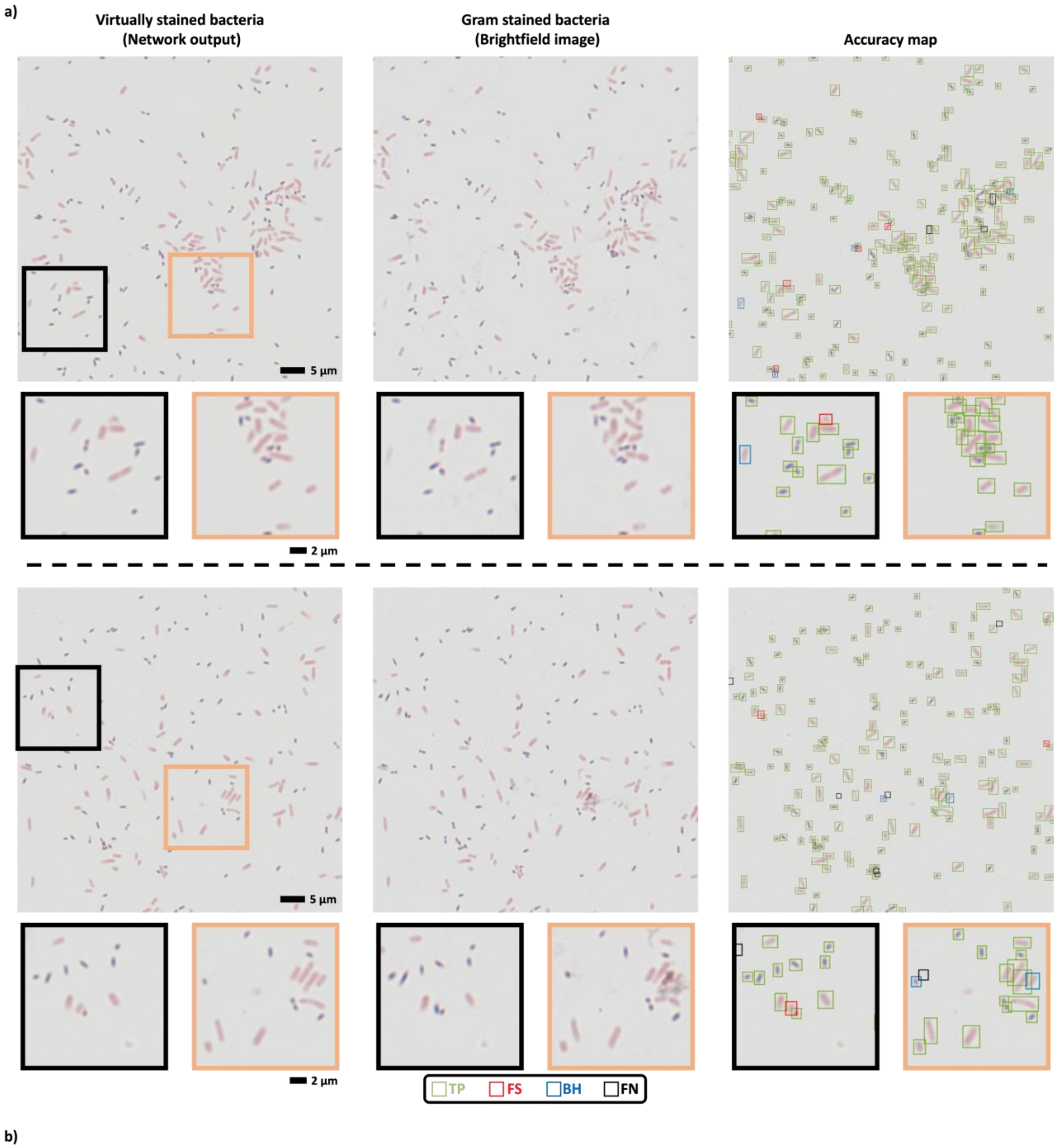

| Precision | Recall | F1-Score | False staining rate | Bacteria hallucination rate | PSNR (dB) | SSIM |
|---|---|---|---|---|---|---|
| 0.9549 | 0.9654 | 0.9602 | 0.0302 | 0.0149 | 38.35 | 0.9685 |

**Figure 2. Comparison of virtual and conventional Gram staining methods on unlabeled bacteria.** (a) Sample FOVs after virtual and standard Gram staining of unlabeled bacteria. Accuracy maps demonstrate successful and failed instances of virtual Gram staining for each individual bacterium in the FOV. $\mathrm{TP}, \mathrm{FS}, \mathrm{BH}$ and $\mathrm{FN}$ represent true positive, false staining, bacterial hallucination, and false negative cases, respectively. (b) Average evaluation metrics computed across all the images of the *mixed test dataset* containing both *E. coli* and *L. innocua*.

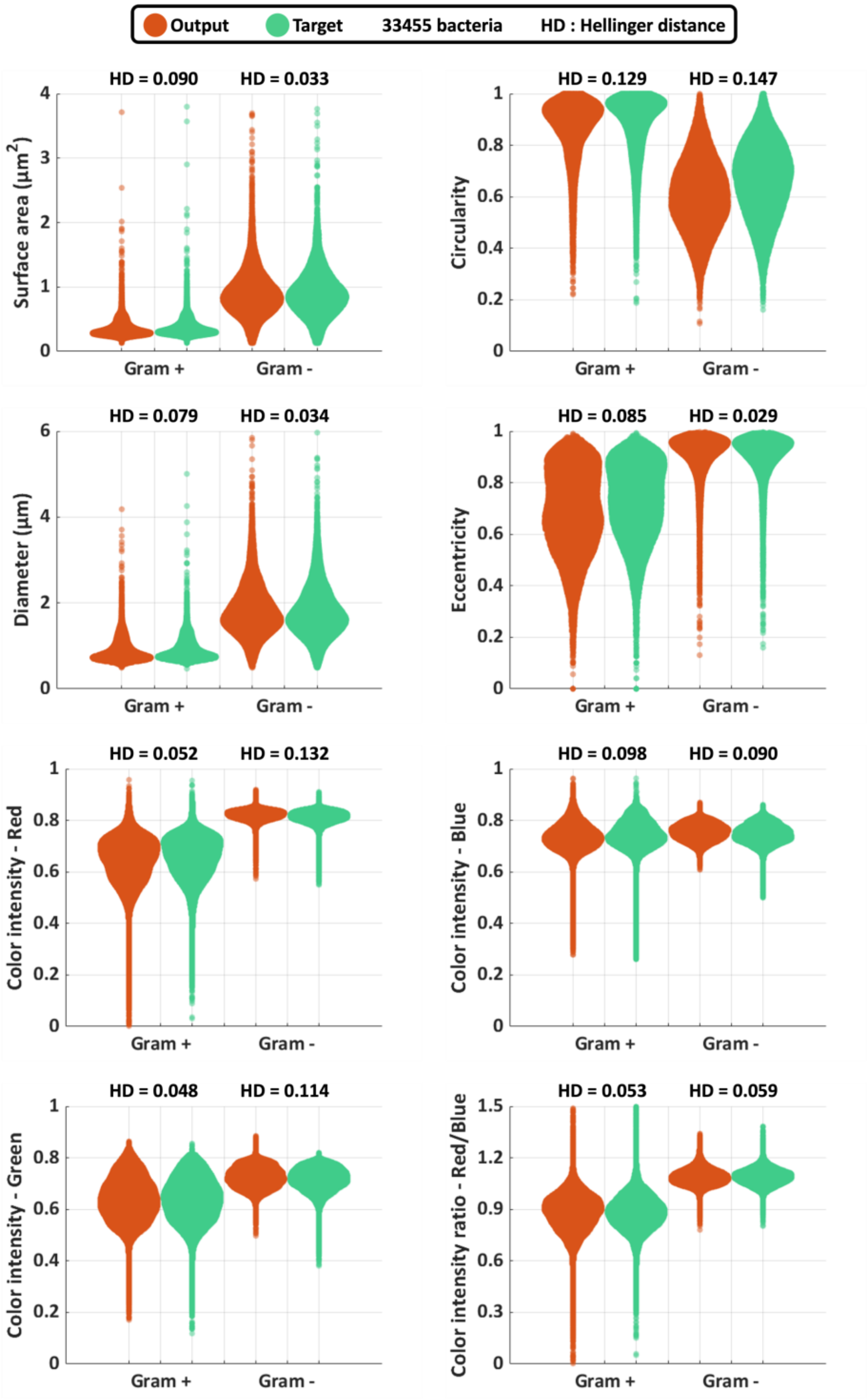

**Figure 3. Comparison of chromatic and morphological features of virtually and chemically stained bacteria.** Feature distributions of segmented bacteria from target (ground truth) and virtually stained output images are illustrated, accompanied by quantitative assessments. Gram-positive (+) and Gram-negative (-) bacteria refer to *L. innocua* and *E. coli*, respectively. Four morphological features including surface area, diameter, circularity, and eccentricity and four chromatic features encompassing color intensities (red, green, and blue) as well as a color intensity ratio (Red/Blue) are visualized. The Hellinger distance (HD) is used to quantify the similarity between the feature distributions of the segmented bacteria from chemically and virtually stained output images.

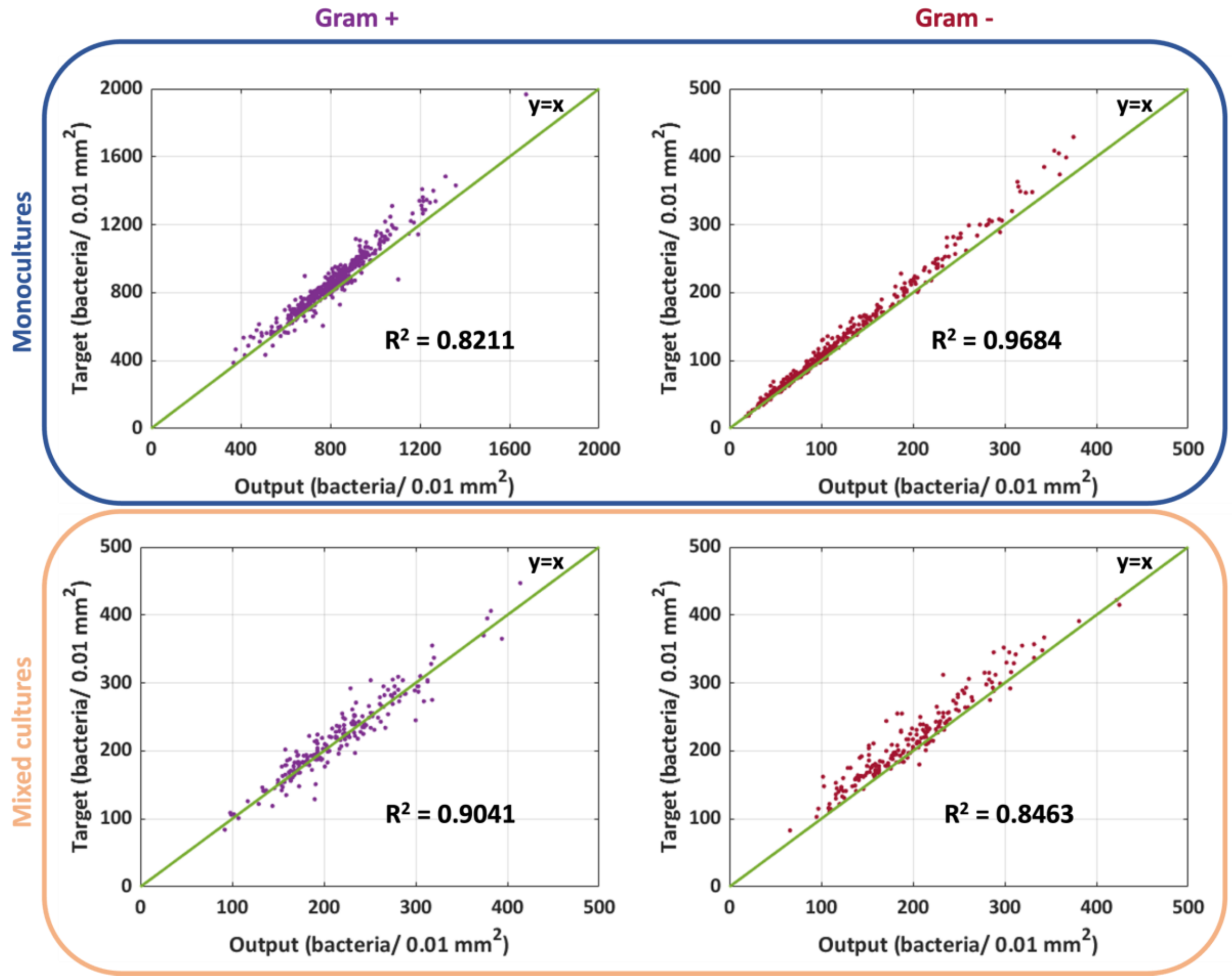

**Figure 4. Analysis of the accuracy of the virtual Gram staining across different bacteria concentration levels.** Predicted and target concentrations of Gram + and Gram – bacteria in monocultures (*E. coli*-only & *L. innocua*-only) and a mixed culture (*E. coli* & *L. innocua*) are visualized. The coefficients of determination ($R^2$ values) for all cases are computed to quantify the goodness-of-fit between the predicted and ground truth bacterial concentrations for each case. Normalized root mean square errors (NRMSEs) for Gram + and Gram - monocultures are 0.082 and 0.102, respectively. In the mixed culture, NRMSEs of Gram + and Gram - cases are 0.080 and 0.111, respectively.

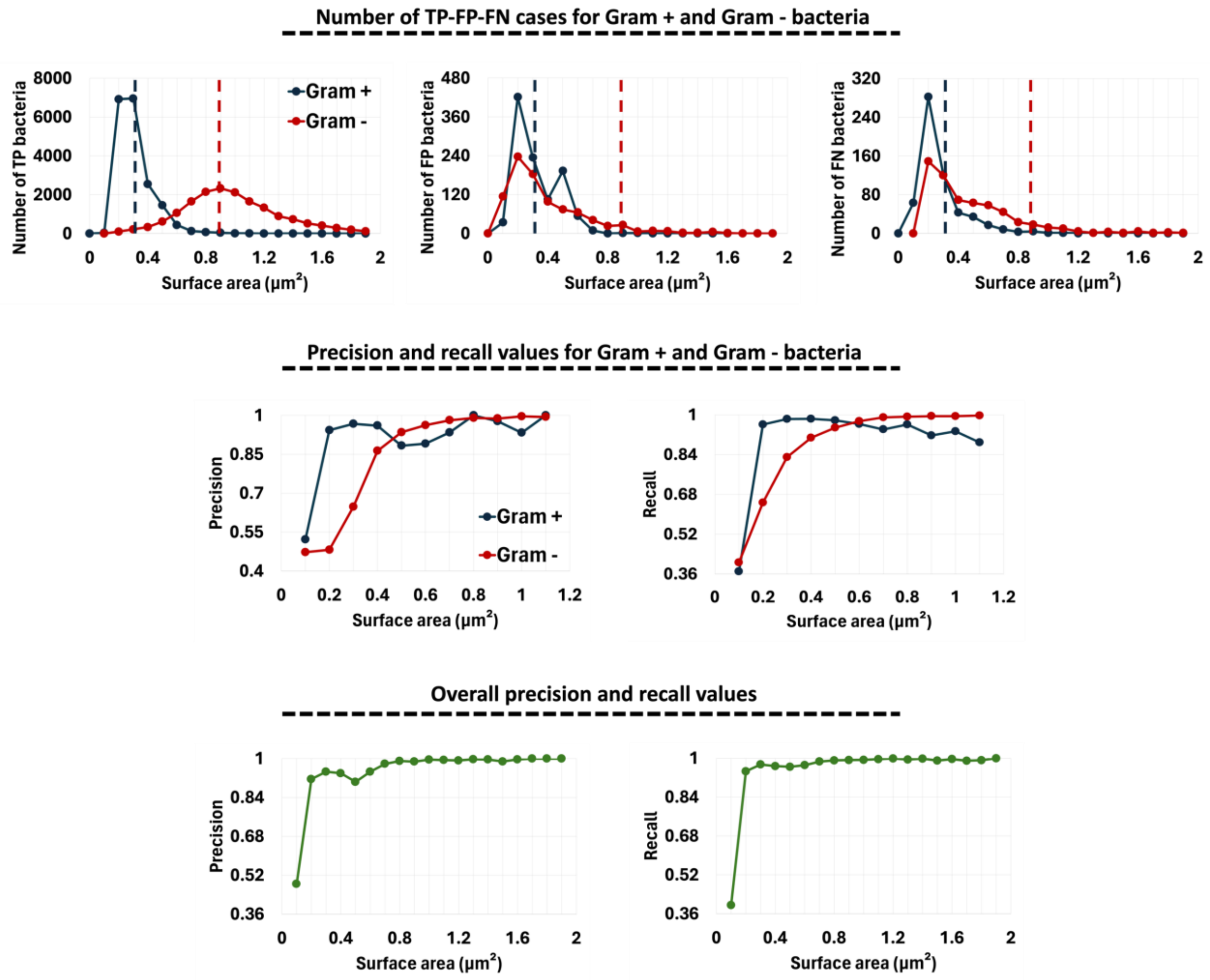

**Figure 5. Accuracy comparison of virtual Gram staining across different size intervals of bacteria.** (Top) Distribution of total $\mathrm{TP}$, $\mathrm{FP}$, and $\mathrm{FN}$ cases across different intervals of bacteria surface area ($\rho \leq \text{surface area} \leq \rho + 0.1\ \mu\mathrm{m}^2$, for each point $\rho$ on the x-axis). (Middle) Precision and recall values of Gram-positive (+, *L. innocua*) and Gram-negative (-, *E. coli*) bacteria are presented for different size intervals. (Bottom) Overall precision and recall values are shown across various bacteria size intervals. The median surface areas of *E. coli* and *L. innocua*, computed using Gram-stained bacteria (ground truth), were 0.892 $\mu m^2$ and 0.317 $\mu m^2$, respectively. Vertical dashed lines represent median surface areas for Gram-positive and Gram-negative bacteria, which were calculated using 5 target images containing Gram-stained bacteria (with minimal overlap of Gram-positive and -negative bacteria).